\documentclass[sn-mathphys,Numbered]{sn-jnl}

\usepackage{graphicx}%
\usepackage{multirow}%
\usepackage{amsmath,amssymb,amsfonts}%
\usepackage{amsthm}%
\usepackage{bm}
\usepackage{braket,comment}
\usepackage{physics}
\usepackage{mathrsfs}%
\usepackage[title]{appendix}%
\usepackage{xcolor}%
\usepackage{textcomp}%
\usepackage{manyfoot}%
\usepackage{booktabs}%
\usepackage{algorithm}%
\usepackage{algorithmicx}%
\usepackage{algpseudocode}%
\usepackage{listings}%
\usepackage{url}
\newcommand{\beq}{\begin{equation}}
\newcommand{\eeq}{\end{equation}}
\newcommand{\beqa}{\begin{eqnarray}}
\newcommand{\eeqa}{\end{eqnarray}}
\newcommand{\bal}{\begin{aligned}[b]}
\newcommand{\eal}{\end{aligned}}
\newcommand*{\bseq}{\begin{subequations}}
\newcommand*{\eseq}{\end{subequations}}
\newcommand{\bpm}{\begin{pmatrix}}
\newcommand{\epm}{\end{pmatrix}}

\theoremstyle{thmstyleone}%
\theoremstyle{thmstyletwo}%

\theoremstyle{thmstylethree}%

\begin{document}

\title[Fokker-Planck equations for a trapped particle in a quantum-thermal Ohmic bath: general theory and applications to Josephson junctions]
{Fokker-Planck equations for a trapped particle in a quantum-thermal Ohmic bath: general theory and applications to Josephson junctions}
\author*[1]{\fnm{Koichiro} \sur{Furutani}}
\email{koichiro.furutani@phd.unipd.it}
\author[1,2]{\fnm{Luca} \sur{Salasnich}}
\affil*[1]{Dipartimento di Fisica e Astronomia "Galileo Galilei" and QTech Center, 
Universit{\`a} di Padova and INFN Sezione di Padova, 
Via Marzolo 8, 35131, Padova, Italy}
\affil[2]{Istituto Nazionale di Ottica del Consiglio Nazionale delle Ricerche,  
Via Nello Carrara 2, 50019, Sesto Fiorentino, Italy}
\date{today}

\abstract{We consider a particle trapped by a generic external potential 
and under the influence of a quantum-thermal Ohmic bath. 
Starting from the Langevin equation, we derive the corresponding Schwinger-Keldysh action. 
Then, within the path-integral formalism, we obtain both the semiclassical Fokker-Planck equation and the quantum Fokker-Planck equation for this out-of-equilibrium system. 
In the case of an external harmonic potential and in the underdamped regime, we find that our Fokker-Planck equations contain an effective temperature $T_{\rm eff}$, which crucially depends on the interplay between quantum and thermal fluctuations in contrast to the classical Fokker-Planck equation. 
In the regime of high temperatures, one recovers the classical Fokker-Planck equation. 
As an application of our result, we also provide the stationary solution of the semiclassical Fokker-Planck equations for a superconducting Josephson circuit and for a Bose Josephson junction, which are experimentally accessible. }

\maketitle

\section{Introduction}

Inspired by the works of Einstein \cite{einstein} and Smoluchowski \cite{smoluchowski} about the Brownian motion of a mesoscopic particle in a fluid, in 1908 Langevin introduced its stochastic equation \cite{langevin1908}. 
Some years later, Johnson \cite{johnson} and Nyquist \cite{nyquist} observed that, in addition to thermal effects, also the quantum mechanical noise plays a relevant role in the electric current of conductors. 
In 1951, the quantum version of the fluctuation-dissipation theorem of Callen and Welton \cite{callen} paved the way to the quantum Langevin equation \cite{koch,ford,metiu}. 
In the first part of this paper, we explicitly show that, from the Langevin equation of a confined particle in a quantum-thermal Ohmic bath \cite{kamenev}, one derives the corresponding semiclassical Schwinger-Keldysh action \cite{schwinger,keldysh}.  
The Schwinger-Keldysh action is usually obtained by adopting a quite different approach, which involves a closed time contour with forward and backward branches in time and where the dynamical variables of the system are doubled to take into account the two branches \cite{schwinger,keldysh,kamenev}. 
Moreover, it is well known that from the Schwinger-Keldysh action one obtains 
in the high-temperature regime the classical Martin-Siggia-Rose action 
\cite{martin}. 
Here, we follow a quite different path: for a particle in contact with a quantum-thermal bath, its semiclassical Schwinger-Keldysh action is obtained directly from the Langevin equation through a Martin-Siggia-Rose action, 
which is indeed the semiclassical Schwinger-Keldysh action functional. 

In the second part, we derive the main results of the paper: 
the Fokker-Planck equations of our specific system. 
These Fokker-Planck equations are partial-differential 
equations describing the probability density $P(q,v,t)$ of finding a particle with position $q$ and velocity $v$ at 
time $t$ \cite{fokker,planck,chang,arnold,jang,collins,pedro} 
in a quantum-thermal Ohmic bath and trapped by a deterministic external 
potential $V(q)$. From the short-time propagator of the transition probability 
associated with the Schwinger-Keldysh action, we obtain both the semiclassical 
and the quantum Fokker-Planck equation for a confined particle 
under the effect of a quantum-thermal Ohmic bath. 
Remarkably, our Fokker-Planck equations are fully analytical 
and contain an effective temperature $T_{\mathrm{eff}}$. 
This effective temperature crucially depends on the interplay 
between quantum fluctuations, characterized by $\hbar\Omega$ with $\hbar$ the reduced Planck constant and $\Omega$ the frequency of harmonic potential, and thermal fluctuations, characterized by the thermal energy 
$k_{\mathrm{B}}T$ with $k_{\mathrm{B}}$ the Boltzmann constant and $T$ the 
temperature. 
Following our approach, we also obtain quantum Fokker-Planck equations for the superconducting phase in a Josephson circuit and for the population imbalance in an atomic Bose Josephson junction, which are described by generalized Langevin equations including quantum and thermal fluctuations.

\section{Langevin equation for a particle in a quantum-thermal 
Ohmic bath}\label{Sec2}

Let us consider a particle of mass $m$ and coordinate $q(t)$ 
under the action of a deterministic 
potential $V(q(t))$ but also of a thermal bath which induces a dissipative 
force $-\gamma {\dot q}(t)$ with damping coefficient $\gamma$ and 
a Gaussian stochastic force $\xi(t)$. 
The stochastic coordinate $q(t)$ of the particle satisfies the equation of motion 
\beq 
m \, {\ddot q}(t) + \gamma \, {\dot q}(t) + {\partial V(q(t))\over \partial q} 
= \xi(t) \; . 
\label{ql}
\eeq
Here, we assume the Markovian dynamics such that the damping term $\gamma\dot{q}(t)$ does not include any memory effect and the equation of motion \eqref{ql} involves only one time variable $t$. 
Given a generic observable $O$ which depends explicitly 
on the Gaussian random variable $\xi(t)$, the stochastic average has the following path integral representation 
\beq 
\langle O \rangle = 
{\int D[\xi(t)] \, O[\xi(t)] \, 
e^{-{1\over 2}\int_{-\infty}^{+\infty}  
\int_{-\infty}^{+\infty} \xi(t) C^{-1}(t-t')\xi(t') \, dt \, dt'} 
\over 
\int D[\xi(t)] \, e^{-{1\over 2}\int_{-\infty}^{+\infty}  
\int_{-\infty}^{+\infty} \xi(t)C^{-1}(t-t')\xi(t') \, dt \, dt'} } ,
\label{delfino}
\eeq
which crucially depends on the choice of the correlation function $C(t)$, 
which, in general, is that $C(t-t')=\langle \xi(t) \xi(t')\rangle$.

Equation \eqref{ql} is called semiclassical 
quantum Langevin equation \cite{koch,ford,metiu} 
provided that the correlation function $C(t)$ is given by 
\beq 
C(t) =\int_{-\infty}^{+\infty}\frac{d\omega}{2\pi} \gamma \,  
\hbar \omega \, \coth\left({\hbar \omega\over 2 k_{\rm B} T}\right) 
\, e^{i\omega \, t}\; , 
\label{cf}
\eeq
where $T$ is the absolute temperature, $k_{\rm B}$ is the Boltzmann constant, 
and $\hbar$ is the reduced Planck constant. The correlation function 
of Eq.~\eqref{cf} is the one of a stochastic quantum-thermal 
Ohmic bath \cite{koch,ford,metiu}. 
The presence of $\gamma$ both in Eqs.~\eqref{ql} and \eqref{cf} 
is a consequence of the fluctuation-dissipation theorem (FDT) \cite{kamenev}. 
Moreover, in Eq.~\eqref{cf} the term $e^{i\omega\, t}$ can be substituted 
by $\cos({\omega\, t})$ because the imaginary part is odd and its integral 
gives zero. 
Equation \eqref{ql} with Eq.~\eqref{cf} 
is called semiclassical because the dynamical variable $q(t)$ 
is not a quantum operator. 
The quantum nature of Eq.~\eqref{ql} is however encoded 
in the correlator $C(t)$ through Eq.~\eqref{cf}. 
Note that in the high-temperature regime $k_{\rm B}T\gg \hbar \omega$, where 
$\coth\left({\hbar \omega/(2 k_{\rm B} T)}\right) \to {2k_{\rm B}T/(\hbar\omega)}$, Eq.~\eqref{cf} gives $C(t) = 2 \gamma \ k_{\rm B} T \ \delta(t)$ with $\delta(t)$ the Dirac delta function and Eq.~\eqref{ql} becomes the 
familiar classical Langevin equation \cite{langevin1908}. 

\section{Martin-Siggia-Rose action}
\label{SecMSR}

To derive the Martin-Siggia-Rose action \cite{martin} (see also 
Refs.~\cite{dominicis,janssen,peliti,aron,tauber,araujo}) from the 
semiclassical quantum Langevin equation \eqref{ql}, we observe that 
the expectation value of the generic observable $O$ can also be written as
\beq
\bal
&\left\langle O\right\rangle=\int D[q(t)]O[q(t)]
\left\langle\delta\left[q-q(t)\right]\right\rangle \\
&=\int D[q(t)]O[q(t)]\left\langle\delta\left[m{\ddot q}(t) + 
\gamma{\dot q}(t) + {\partial V(q(t))\over \partial q}-
\xi(t)\right]\right\rangle ,
\eal
\eeq
where the Dirac delta function $\delta(x)$ appears because one considers the path integral over all possible $q(t)$ but with the constraint that $q(t)$ satisfies Eq.~\eqref{ql}. This constraint ensures the correct implicit dependence of $q(t)$ with respect to $\xi(t)$. 
Taking into account the path integral representation of $\delta(x)$, we have 
\beq
\bal
\left\langle O\right\rangle&=\int D[q(t),\Tilde{q}(t)]O[q(t)]
\left\langle \exp\left[i\int^{\infty}_{-\infty} dt\Tilde{q}(t)
\left[m\ddot{q}(t)+\gamma\dot{q}(t)+\frac{\partial V(q(t))}{\partial q}-
\xi(t)\right]\right]\right\rangle ,
\eal
\eeq
where ${\tilde q}(t)$ is an auxiliary response field. 
We now use Eq.~\eqref{delfino} and the properties of Gaussian integrals obtaining 
\beq
\bal
\left\langle O\right\rangle&=\int D[q(t),\Tilde{q}(t)]O[q(t)]
e^{iS\left[q(t),\Tilde{q}(t)\right]/\hbar} ,
\eal
\eeq
where 
\beq
\bal
S[q(t),{\tilde q}(t)]/\hbar &= -\int_{-\infty}^{+\infty} {\tilde q}(t) \left[ 
m \, {\ddot q}(t) + \gamma \, {\dot q}(t) + 
{\partial V(q(t))\over \partial q} \right] \, dt \\ 
&+\frac{i}{2} \int_{-\infty}^{+\infty}
\int_{-\infty}^{+\infty} {\tilde q}(t) C(t-t') {\tilde q}(t') \, dt \, dt' ,
\eal
\label{msr}
\eeq
remembering that the stochastic noise $\xi(t)$ 
is time-translation invariant, namely $C(t-t')=\langle \xi(t) \xi(t') \rangle = 
\langle \xi(t-t') \xi(0)\rangle$. 
The functional of Eq.~\eqref{msr}, with $C(t)=\langle \xi(t) \xi(0)\rangle$ 
given by Eq.~\eqref{cf}, is our Martin-Siggia-Rose action \cite{martin}. 

We have seen that in the high-temperature regime $C(t)=2\gamma k_{\rm B} T \delta(t)$. 
In this regime, Eq.~\eqref{msr} is much simpler and the action is known as the 
classical Martin-Siggia-Rose action \cite{martin}. 
Quite remarkably, in this classical high-temperature regime, one can easily 
perform the path integral over the dual variable ${\tilde q}(t)$, 
obtaining 
\beq 
\left\langle O\right\rangle = \int D[q(t)] \ O[q(t)] \ 
e^{- S_{\rm OM}[q(t)]} ,
\eeq 
with the real effective action 
\beq 
S_{\rm OM}[q(t)] = {1\over 4\gamma k_{\rm B} T}
\int_{-\infty}^{+\infty} \left[ m \, {\ddot q}(t) + \gamma \, {\dot q}(t) + 
{\partial V(q(t))\over \partial q} \right]^2 \, dt ,  
\label{aom}
\eeq
that is called Onsager-Machlup \cite{onsager} action according to Olender and Elber \cite{olender}. 

\section{Semiclassical Schwinger-Keldysh action}
\label{SK}

Let us restrict ourselves to a harmonic potential $V(q)=m\Omega^{2}q^{2}/2$. 
In the underdamped regime $\gamma\ll m\Omega^{2}/\omega_{\rm cut}$ with $\omega_{\rm cut}$ being the ultraviolet cutoff frequency associated with the zero-point fluctuations in Eq.~\eqref{cf}, we apply the approximation \cite{huard2006,huard2007}
\beq 
C(t)\simeq \gamma\hbar\Omega\coth{\left(\dfrac{\hbar\Omega}{2k_{\rm B}T}\right)}\delta(t)
=2\gamma k_{\rm B}T_{\rm eff}\delta(t) ,
\label{whiteapp}
\eeq
which is white noise including quantum fluctuations. 
The effective temperature is defined by
\beq
T_{\rm eff}=\frac{\hbar\Omega}{2k_{\rm B}}\coth{\left(\frac{\hbar\Omega}{2k_{\rm B}T}\right)} .
\label{Teff}
\eeq
This approximation is justified for the following reason. 
The Langevin equation \eqref{ql} in the long-time limit gives \cite{furutani}
\beq
\Tilde{q}(\omega)=\int^{\infty}_{-\infty}dt\, q(t)e^{-i\omega t}
=\Tilde{\chi}(\omega)\Tilde{\xi}(\omega),
\label{resp}
\eeq
where $\Tilde{\xi}(\omega)=\int^{\infty}_{-\infty}dt\xi(t)e^{-i\omega t}$ and
\beq
\Tilde{\chi}(\omega)=\dfrac{1}{-m\omega^{2}-i\gamma\omega+m\Omega^{2}} ,
\label{chi}
\eeq
is the response function in the frequency domain. 
Equation \eqref{resp} leads to the correlation function 
\beq
\langle q(t\to\infty)^{2}\rangle
=\int^{\infty}_{-\infty}\frac{d\omega}{2\pi}\Tilde{C}(\omega)\abs{\Tilde{\chi}(\omega)}^{2} ,
\label{phi2chi}
\eeq
with $\Tilde{C}(\omega)=\int^{\infty}_{-\infty}dte^{-i\omega t}C(t)=\gamma\hbar\omega\coth{[\hbar\omega/(2k_{\rm B}T)]}$. 
The autocorrelation function of the conjugate momentum also involves the combination of $\Tilde{C}(\omega)\abs{\Tilde{\chi}(\omega)}^{2}$. 
In the underdamped limit $\gamma\ll m\Omega^{2}/\omega_{\rm cut}$, the response function \eqref{chi} is dominant only around $\omega=\pm\Omega$ as
\beq
\gamma\abs{\omega}\abs{\Tilde{\chi}(\omega)}^{2}
\to\dfrac{\pi}{2m^{2}\Omega}\left[\delta(\omega-\Omega)+\delta(\omega+\Omega)\right] .
\eeq
Consequently, in Eq.~\eqref{phi2chi}, we can safely use approximation that $\Tilde{C}(\omega)\simeq\Tilde{C}(\Omega)$, which justifies the white noise approximation in Eq.~\eqref{whiteapp} in the underdamped limit $\gamma\ll m\Omega^{2}/\omega_{\rm cut}$. 
In the classical limit $\hbar\Omega\ll k_{\rm B}T$, we recover $T_{\rm eff}=T$. 
In the low-temperature regime $\hbar\Omega\gg k_{\rm B}T$, on the other hand, Eq.~\eqref{whiteapp} provides $T_{\rm eff}=\hbar\Omega/(2k_{\rm B})$. 

By using Eq.~\eqref{whiteapp}, the action \eqref{msr} with the kernel reads  
\beq
\bal
S[q(t),{\tilde q}(t)]/\hbar &=
-\int_{-\infty}^{+\infty} {\tilde q}(t) 
\left[ m \, {\ddot q}(t) + \gamma \, {\dot q}(t) + 
m\Omega^{2}q \right] \, dt  \\
&+ \frac{i\gamma}{2} \int_{-\infty}^{+\infty} 
2k_{\rm B}T_{\rm eff} {\tilde q}(t)^2 \, dt .
\eal
\label{kel}
\eeq  
Remarkably, Eq.~\eqref{kel} is very similar to the Schwinger-Keldysh action 
\cite{schwinger,keldysh} of a quantum particle in contact with an Ohmic bath. The only difference is due to the fact that instead of $\hbar{\tilde q}\, \partial V(q)/\partial q$ in the exact Schwinger-Keldysh action there is $[V(q+\hbar{\tilde q})-V(q-\hbar{\tilde q})]/2$. 
See, for instance, page.~33 of Ref.~\cite{kamenev}. 
Clearly, $[V(q+\hbar{\tilde q})-V(q-\hbar{\tilde q})]/2\simeq 
\hbar{\tilde q}\, \partial V(q)/\partial q$ under the assumption of a small $\hbar{\tilde q}$. 
It is important to stress that the Echern-Schon-Ambegaokar action \cite{eckern} used for superconducting Josephson junctions is nothing else than the exact Schwinger-Keldysh action. 
At zero temperature, the Schwinger-Keldysh action has been used to study the effect of quantum noise in the quantum phase transition of a Josephson junction \cite{altman}. 
In the classical limit $T_{\rm eff}\to T$, one can readily find that Eq.~\eqref{kel} coincides with the classical dissipative action \cite{kamenev}. 

We call Eq.~\eqref{kel} semiclassical Schwinger-Keldysh action. 
Indeed, the classical Schwinger-Keldysh action, which is 
nothing else than the classical Martin-Siggia-Rose action, 
is obtained with $T_{\rm eff}\to T$. 
Then, by functional integrating over ${\tilde q}(t)$ one recovers again the 
Onsager-Machlup action \eqref{aom}.

\section{Fokker-Planck equations}\label{SecFP}

In this section, we derive the corresponding Fokker-Planck equation for 
the MSR action in Eq.~\eqref{msr} or the Schwinger-Keldysh action 
in Eq.~\eqref{kel}. 

\subsection{Semiclassical Fokker-Planck equation}

Let us rewrite the action of Eq.~\eqref{msr} as
\beq
e^{iS[q(t),\Tilde{q}(t)]/\hbar}\equiv
\int D[v(t),\lambda(t)] e^{iS_{\mathrm{e}}[q(t),v(t),\Tilde{q}(t),
\lambda(t)]/\hbar} ,
\eeq
\beq
\bal
S_{\mathrm{e}}[q(t),v(t),\Tilde{q}(t),\lambda(t)] &=\int^{\infty}_{-\infty}dt 
L_{\mathrm{e}}[q(t),v(t),\Tilde{q}(t),\lambda(t)] \\
&=\int^{\infty}_{-\infty} dt \hbar\Bigg[-m\Tilde{q}(t)\dot{v}(t)-m\Tilde{q}(t)
\frac{\partial F(q,v)}{\partial v} \\
&+\frac{i}{2}\int^{\infty}_{-\infty} 
dt'\Tilde{q}(t)C(t-t')\Tilde{q}(t')+\lambda(t)\left[\dot{q}(t)-v(t)\right]\Bigg] ,
\eal
\label{Se}
\eeq
where $S_{\mathrm{e}}[q(t),v(t),\Tilde{q}(t),\lambda(t)]$ is a new effective 
action with a velocity field $v(t)$ and an auxiliary field $\lambda(t)$ 
as a Lagrange multiplier that guarantees $v(t)=\dot{q}(t)$ by 
$\delta S_{\mathrm{e}}/\delta\lambda=0$ \cite{das}, and
\beq
F(q,v)\equiv \frac{\gamma}{2m}v^{2}+\frac{V'(q)}{m}v .
\eeq

\begin{figure}[t]
\centering
\includegraphics[width=0.6\textwidth]{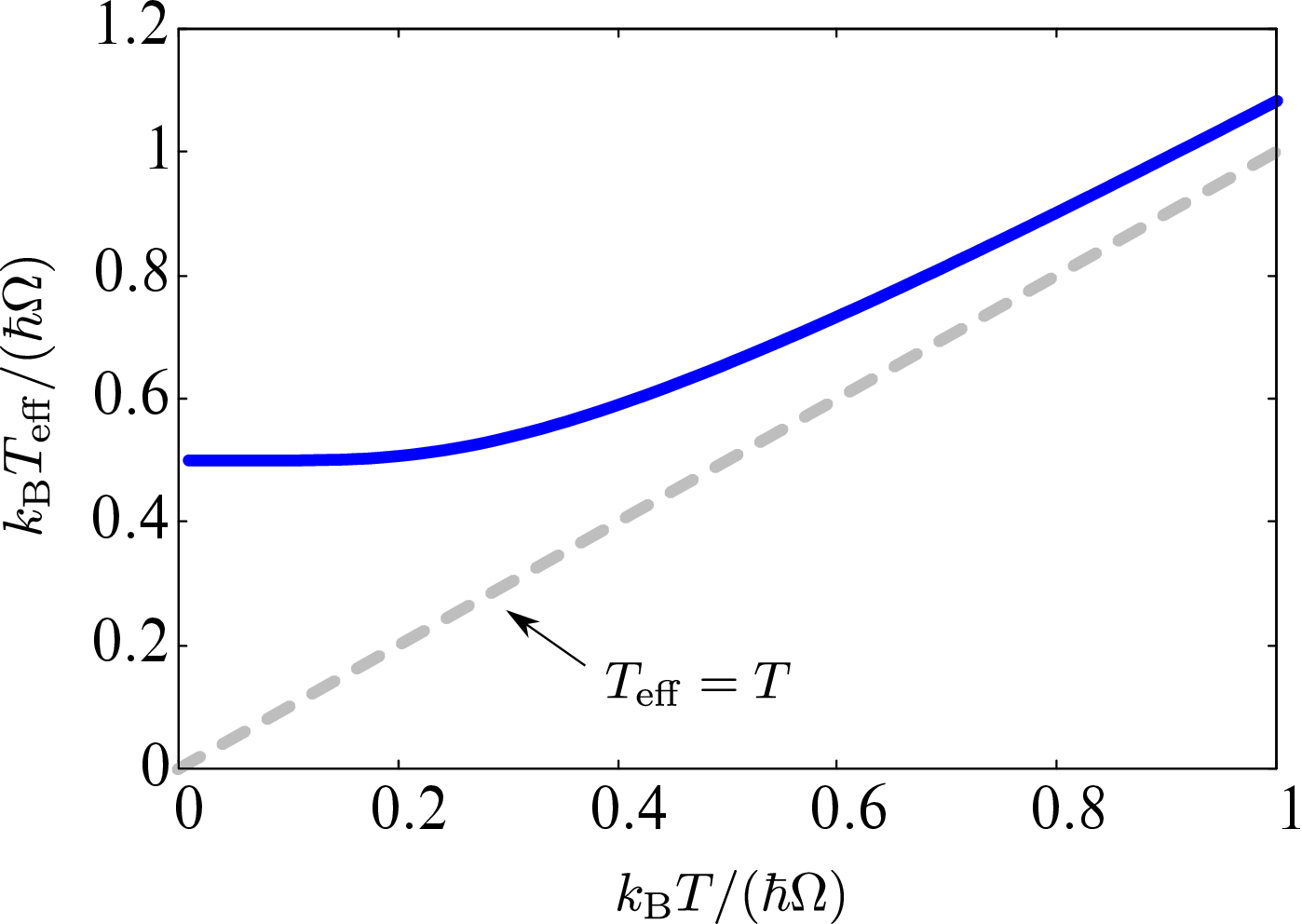}
\caption{Effective temperature in Eq.~\eqref{Teff}. 
The dashed line stands for $T_{\mathrm{eff}}=T$. }
\label{Tefffig}
\end{figure}

The effective Lagrangian $L_{\mathrm{e}}[q(\bar{t}),v(\bar{t}),
\Tilde{q}(\bar{t}),\lambda(\bar{t})]$ 
can be used to introduce the propagator \cite{das}
\beq
\bal
K(q',v',t'|q,v,t) &\equiv \int_{q(t)=q}^{q(t')=q'} 
D[q(\bar{t})] \int_{v(t)=v}^{v(t')=v'} D[v(\bar{t})] 
 \\
&\times \int D[\Tilde{q}(\bar{t}),\lambda(\bar{t})]e^{i\int^{t'}_{t}d\bar{t} L_{\mathrm{e}}[q(\bar{t}),v(\bar{t}),
\Tilde{q}(\bar{t}),\lambda(\bar{t})]/\hbar} \; ,
\eal
\label{fis07}
\eeq
which gives the transition probability from the initial configuration 
$(q,v,t)$ to the final configuration $(q',v',t')$. 
It follows that the probability $\mathcal{P}(q,v,t)$ of finding 
the system in the configuration $(q,v,t)$ satisfies the 
convolution equation
\beq
\mathcal{P}(q',v',t') = 
\int_{-\infty}^{+\infty} dq \int_{-\infty}^{+\infty} dv \, 
K(q',v',t'|q,v,t) \, \mathcal{P}(q,v,t) \; . 
\label{fis03}
\eeq
In Appendix \ref{AppA}, we show how to derive the semiclassical 
Fokker-Planck equation from this expression taking into account 
Eq.~\eqref{fis07}. The final result of this derivation is \cite{das,kleinert}
\beq
\partial_{t}\mathcal{P}(q,v,t) 
=\left[-v\partial_{q}+ \frac{\gamma}{m}
\partial_{v}v 
+ \frac{V'(q)}{m}\partial_{v}
+\frac{\gamma k_{\mathrm{B}} T_{\mathrm{eff}}}{m^{2}} 
\partial_{v}^2\right]\mathcal{P}(q,v,t) \; , 
\label{FPeq}
\eeq
under Eq.~\eqref{whiteapp} where $V(q)$ is the harmonic potential of frequency $\Omega$. 
In the classical limit $\hbar\to0$, or equivalently $k_{\mathrm{B}}T\gg \hbar\Omega$,  
Eq.~\eqref{FPeq} reduces to 
the familiar classical one \cite{kamenev,hashitsume,kubo}
\beq
\partial_{t}\mathcal{P}(q,v,t) 
=\left[-v\partial_{q}+ 
\frac{\gamma}{m}\partial_{v}v + \frac{V'(q)}{m}\partial_{v}
+\frac{\gamma k_{\mathrm{B}}T}{m^{2}}
\partial_{v}^2\right]\mathcal{P}(q,v,t) \; . 
\label{clFPeq}
\eeq
On the other hand, at $T=0$, the effective temperature has a finite 
minimum $T_{\mathrm{eff}}=\hbar\Omega/\left(2k_{\mathrm{B}}
\right)$ as shown in Fig.~\ref{Tefffig}. 
Figure \ref{Tefffig} illustrates that the difference from the original 
temperature is significant in the low-temperature regime due to the quantum 
effects. 

Quite remarkably, the semiclassical Fokker-Planck equation \eqref{FPeq} 
has the following stationary analytical solution 
\beq
\mathcal{P}_{\text{stat}}(q,v)=Z^{-1}e^{-\left[mv^{2}/2+V(q)\right]/
\left(k_{\mathrm{B}}T_{\mathrm{eff}}\right)} \; ,
\label{Pstat}
\eeq
where 
\beq 
Z\equiv\int^{\infty}_{-\infty}dq\int^{\infty}_{-\infty}
dv \, e^{-\left[mv^{2}/2+V(q)\right]/\left(k_{\mathrm{B}}
T_{\mathrm{eff}}\right)} \; . 
\eeq
With a harmonic potential $V(q)=m\Omega^{2}q^{2}/2$, the stationary solution gives the second moments of the coordinate and velocity as 
\beq
\langle q^{2}\rangle=\int^{\infty}_{-\infty}dq\int^{\infty}_{-\infty}dv q^{2}\mathcal{P}_{\rm stat}(q,v)
=\frac{k_{\rm B}T_{\rm eff}}{m\Omega^{2}},
\label{qvar}
\eeq
and
\beq
\langle v^{2}\rangle=\int^{\infty}_{-\infty}dq\int^{\infty}_{-\infty}dv v^{2}\mathcal{P}_{\rm stat}(q,v)
=\frac{k_{\rm B}T_{\rm eff}}{m}.
\label{vvar}
\eeq
These second moments are exactly consistent with a Langevin analysis in the underdamped limit \cite{brabert,furutani}. 
In the classical limit $T_{\rm eff}\to T$, Eqs.~\eqref{qvar} and \eqref{vvar} recover the equipartition of energy. 
Due to the quantum correction in Eq.~\eqref{Teff}, both of Eqs.~\eqref{qvar} and \eqref{vvar} are proportional to $\hbar\Omega$ in the low-temperature regime. 

Note that Eqs.~\eqref{qvar} and \eqref{vvar} are the underdamped results which are, strictly speaking, justified in the underdamped limit $\gamma\ll m\Omega^{2}/\omega_{\rm cut}$. 
With finite damping constant $\gamma>m\Omega^{2}/\omega_{\rm cut}$, the second moments of the velocity should have a logarithmic ultraviolet divergence \cite{brabert}. 
We stress that, however, we are restricting ourselves to the underdamped limit and do not consider such cases throughout this paper. 
Only in this underdamped limit, do the second moments of the coordinate or the momentum have clear physical interpretation as they recover the equipartition of energy in the classical limit \cite{brabert,furutani}. 

\subsection{Quantum Fokker-Planck equation}

As mentioned in Sec.~\ref{SK}, the dissipative Schwinger-Keldysh action is 
different from the MSR action in terms of the potential term in the higher 
order of $\Tilde{q}$. Let us call the corresponding Fokker-Planck equation 
derived from the Schwinger-Keldysh action the quantum Fokker-Planck equation 
since it includes higher order of the quantum component $\Tilde{q}$, which 
should coincide with the semiclassical one to the first order of $\Tilde{q}$. 

To obtain the quantum Fokker-Planck equation, we can simply substitute the 
potential term in Eq.~\eqref{Se} as
\beq
\hbar\Tilde{q}V'(q) 
\to \frac{V(q+\hbar\Tilde{q})-V(q-\hbar\Tilde{q})}{2} 
=\sum_{n=0}^{\infty}\left(\hbar\Tilde{q}\right)^{2n+1}\frac{V^{(2n+1)}(q)}{(2n+1)!} \; ,
\eeq
where $V^{(k)}(q)$ denotes the $k$-derivative of $V(q)$ in $q$. Since it is 
time-local, one can proceed to the Fokker-Planck equation in a similar manner 
as in the last section. Eventually, one obtains the quantum Fokker-Planck 
equation
\beq
\partial_{t}\mathcal{P}(q,v,t) 
=\left[-v\partial_{q}+
{\gamma\over m}\partial_{v}v + \frac{\gamma k_{\mathrm{B}} T_{\mathrm{eff}}}{m^{2}}
 \partial_{v}^2 
+\partial_{v} \left(\partial_{v}\Tilde{F}\right) \right] \mathcal{P}(q,v,t) \; ,
\label{qFPeq}
\eeq
where $\Tilde{F}$, that gives the difference from semiclassical equation 
\eqref{FPeq}, is defined as
\beq
\bal
\Tilde{F}(q,v) &\equiv \frac{v}{m}\sum_{n=0}^{\infty}
\left(\frac{\hbar\partial_{v}}{im}\right)^{2n}\frac{V^{(2n+1)}(q)}{(2n+1)!}
 \\
&=\frac{V'(q)}{m}v+\frac{v}{m}\sum_{n=1}^{\infty}\left(\frac{\hbar\partial_{v}}{im}\right)^{2n}
\frac{V^{(2n+1)}(q)}{(2n+1)!} \\
&=\Omega^{2}qv .
\eal
\label{qFtilde}
\eeq
For a harmonic potential $V(q)=m\Omega^{2}q^{2}/2$, as in the last equality in Eq.~\eqref{qFtilde}, the higher-order derivatives in Eq.~\eqref{qFtilde} vanish. 
With a generic potential, the presence of infinite derivatives makes it quite difficult to solve it in full generality. 
While a truncation of the series with $n=1$ or $n=1,2$ could be used to find reliable corrections to the semiclassical result, it may lead to negative probability distribution \cite{pawula}. 

\section{Fokker-Planck equation for Josephson junctions}\label{SecFPJJ}

In a resistively and capacitively shunted Josephson (RCSJ) junction, the superconducting phase $\phi(t)$ obeys the generalized Langevin equation \cite{brandt,furutani}
\beq
\frac{\hbar^{2}}{2E_{C}}\ddot{\phi}+\frac{\hbar\alpha}{2\pi}\dot{\phi}+E_{J}\sin{\phi}-\frac{\Phi_{0}I_{\mathrm{ext}}}{2\pi}=\xi ,
\eeq
where $E_{C}=(2e)^{2}/(2C)$ is the charging energy with a capacitance $C$, $\alpha=R_{Q}/R$ is the ratio between the critical resistance $R_{Q}=h/(2e)^{2}$ and the resistance $R$, $E_{J}$ is the Josephson energy, $I_{\mathrm{ext}}$ is the external current, and $\Phi_{0}=h/(2e)$ is the magnetic flux quantum. 
The current noise $\xi(t)$ originates from the shunted resistor and is assumed to satisfy the FDT in Eq.~\eqref{cf}. 
It has been theoretically predicted that the system is superconducting below the critical resistance $\alpha>1$ while it is insulating above the resistance $\alpha<1$ \cite{schmid,bulgadaev}. 
The ultraviolet cutoff frequency can be chosen as $\omega_{\rm cut}=\Delta/\hbar$ with $\Delta$ being the superconducting gap. 
The approximation \eqref{whiteapp} is valid within the underdamped limit $\alpha\ll 2\pi E_{J}/\Delta$ \cite{furutani}.  

\begin{figure*}
\centering
\includegraphics[width=0.49\textwidth]{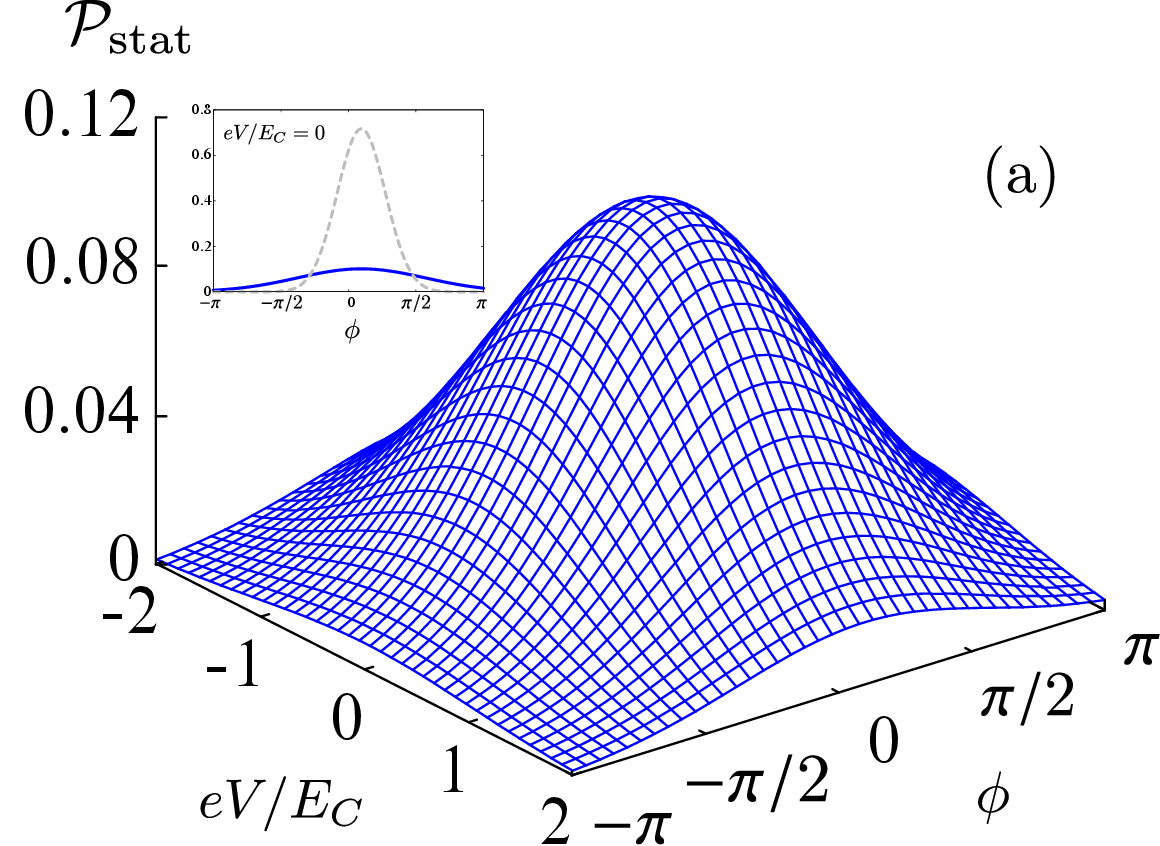}
\includegraphics[width=0.49\textwidth]{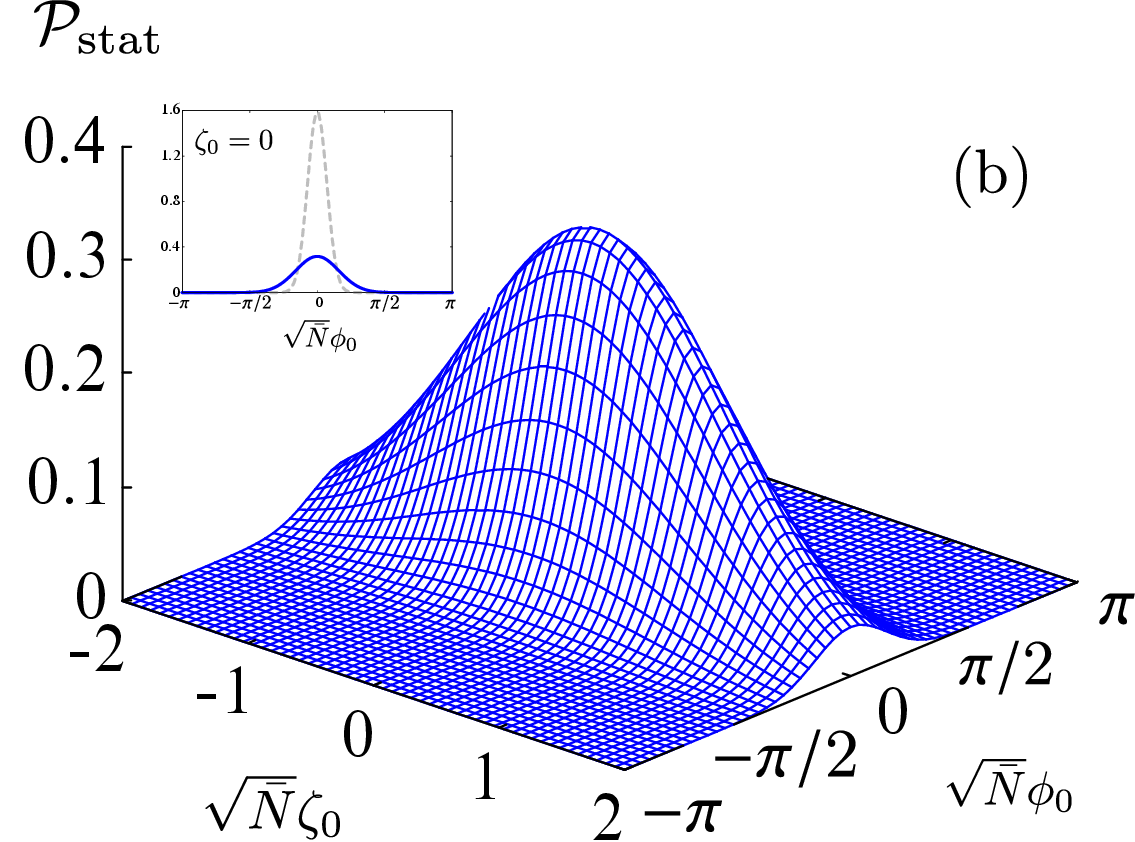}
\caption{Stationary solutions of the semiclassical Fokker-Planck equations of a RCSJ junction (a) and BJJ (b). 
The left panel (a) illustrates the result of Eq.~\eqref{RCSJFPeqstatcl} for $E_{J}/E_{C}=1$, $k_{\mathrm{B}}T/E_{C}=0.1$, and $\Phi_{0}I_{\text{ext}}/(2\pi E_{J})=0.3$.
The right panel (b) shows the result of Eq.~\eqref{BJJFPeqstat} for $\Omega/\Omega_{\text{R}}=0.5$ and $k_{\mathrm{B}}T/(\hbar\Omega)=0.1$. 
The gray dashed curves in the insets stand for the probability in the classical limit $T_{\rm eff}\to T$. }
\label{PstatJJ}
\end{figure*}

We can identify this Langevin equation as Eq.~\eqref{ql} by replacing $q(t)\to\phi(t)$, $v\to\dot{\phi}=2\pi V/\Phi_{0}$, $m\to\hbar^{2}/(2E_{C})$, $\gamma\to\hbar\alpha/(2\pi)$, and $V(q)\to U_{\mathrm{wash}}[\phi]\equiv -E_{J}\cos{\phi}-\Phi_{0}I_{\mathrm{ext}}/(2\pi)\cdot\phi$. 
Consequently, following the procedures in the last sections, we obtain the quantum Fokker-Planck equation for the RCSJ junction as
\beq
\bal
\partial_{t}\mathcal{P}(\phi,V,t)
&=\Bigg[-\frac{2\pi V}{\Phi_{0}}\partial_{\phi}+\frac{\alpha E_{C}}{\pi\hbar}\partial_{V}V
+\frac{\alpha\Phi_{0}^{2}E_{C}^{2}}{2\pi^{2}\hbar^{3}}k_{\mathrm{B}}T_{\mathrm{eff}}\partial_{V}^{2} \\
&+\left(\frac{\Phi_{0}}{2\pi}\right)^{2}\partial_{V}\left(\partial_{V}\Tilde{F}[\phi,V]\right)\Bigg]\mathcal{P}(\phi,V,t),
\eal
\label{RCSJFPeq}
\eeq
with $V(t)=\Phi_{0}\dot{\phi}/(2\pi)$ the voltage. 
The function $\Tilde{F}[\phi,V] $ is given by
\beq
\bal
\Tilde{F}[\phi,V] 
&=\frac{4\pi E_{C}}{\Phi_{0}\hbar^{2}}\Bigg[\frac{\partial U_{\text{wash}}[\phi]}{\partial\phi} 
+\sum_{n=1}^{\infty}\left(\frac{\Phi_{0}E_{C}}{i\pi\hbar}\partial_{V}\right)^{2n}\frac{U^{(2n+1)}_{\mathrm{wash}}[\phi]}{(2n+1)!}\Bigg]V \\
&=\frac{4\pi E_{C}}{\Phi_{0}\hbar^{2}}\left(-\dfrac{\Phi_{0}I_{\rm ext}}{2\pi}+E_{J}\phi\right)V ,
\eal
\label{FRCSJ}
\eeq
where, in the last row, we used $U_{\rm wash}[\phi]\simeq-E_{J} -\Phi_{0}I_{\rm ext}/(2\pi)\cdot\phi+E_{J}\phi^{2}/2$. 
We can find the semiclassical stationary solution as
\beq
\mathcal{P}^{\text{cl}}_{\text{stat}}(\phi,V)=Z^{-1}\mathrm{exp}\left[-\frac{E_{C}}{\pi k_{\mathrm{B}}T_{\mathrm{eff}}}\left[\left(\frac{eV}{E_{C}}\right)^{2}+\frac{U_{\text{wash}}[\phi]}{E_{C}}\right]\right] ,
\label{RCSJFPeqstatcl}
\eeq
which is illustrated in Fig.~\ref{PstatJJ}(a) for $E_{J}/E_{C}=1$, $k_{\mathrm{B}}T/E_{C}=0.1$, and $\Phi_{0}I_{\text{ext}}/(2\pi E_{J})=0.3$. 
Also in Fig.~\ref{PstatJJ}(a), we use $U_{\text{wash}}[\phi]\simeq-E_{J}-\Phi_{0}I_{\text{ext}}/(2\pi)\cdot\phi+E_{J}\phi^{2}/2$. 
The effective temperature is given by Eq.~\eqref{Teff} with $\Omega=(2E_{C}E_{J})^{1/2}/\hbar$. 
Equation \eqref{RCSJFPeqstatcl} indicates that, in the stationary configuration, the voltage $V$ is more localized as one increases $E_{C}/(k_{\mathrm{B}}T_{\mathrm{eff}})$, and the superconducting phase $\phi$ is localized with a large $E_{J}/E_{C}$. 
Experimentally, one can observe $E_{J}/E_{C}\simeq3.8\times10^{5}$ in a RCSJ circuit \cite{devoret}, which reflects the highly localized phase in the superconducting circuit. 

In a one-dimensional Bose Josephson junction (BJJ) in a head-to-tail configuration, we have two one-dimensional Bose gasses in contact through a tunnel coupling $J(x)=J_{0}L\delta(x)$ at a point $x=0$ where $J_{0}$ is the strength of the Josephson coupling and $L$ is the system size. 
The zero-mode of the population imbalance $\zeta_{0}$ also obeys the Langevin equation \cite{minguzzi,binanti}
\beq
\ddot{\zeta}_{0}+\gamma\dot{\zeta}_{0}+\Omega^{2}\zeta_{0}=\frac{\sqrt{M}\Omega}{\hbar\bar{\rho}}\xi ,
\label{BJJLangevin}
\eeq
in the linear regime $\abs{\zeta_{0}(t)}\ll1$ with $\Omega$ the Josephson frequency, $\gamma$ the damping constant associated with the Josephson coupling $J_{0}$, $M$ the effective mass related to the interparticle interaction strength $g$, and $\bar{\rho}$ the average atomic density. 
The stochastic noise $\xi(t)$ satisfies the FDT in Eq.~\eqref{cf}. 
The ultraviolet cutoff can be chosen as $\omega_{\rm cut}=2\pi c\bar{\rho}\simeq10^{4}\Omega$ where $c=(g\bar{\rho}/m)^{1/2}$ is the speed of sound and $m$ is the atomic mass \cite{binanti,pigneur}. 
Then, the approximation \eqref{whiteapp} is valid if $\bar{N}J_{0}/Mc^{2}\ll10^{-8}$ with $\bar{N}\equiv\bar{\rho}L$ being the average number of atoms, which is the Josephson regime in which the tunneling energy $\bar{N}J_{0}$ is much smaller than the kinetic energy $Mc^{2}$. 
For this one-dimensional BJJ, we find the quantum Fokker-Planck equation as
\beq
\partial_{t}\mathcal{P}(\zeta_{0},\phi_{0},t)
=\Bigg[\Omega_{\text{R}}\phi_{0}\partial_{\zeta_{0}}+
\gamma\partial_{\phi_{0}}\phi_{0} + \frac{2g}{\hbar^{2}c}
k_{\mathrm{B}} T_{\mathrm{eff}} \partial_{\phi_{0}}^2 
+\frac{1}{\Omega^{2}_{\text{R}}}\partial_{\phi_{0}} \left(\partial_{\phi_{0}}\Tilde{F}\right) \Bigg] \mathcal{P}(\zeta_{0},\phi_{0},t) \; ,
\label{BJJFPeq}
\eeq
with $\Omega_{\text{R}}\equiv J_{0}/\hbar$ the Rabi frequency and $\phi_{0}(t)=-\dot{\zeta}_{0}(t)/\Omega_{\text{R}}$ the zero-mode of the relative phase. 
The function $\Tilde{F}(\zeta_{0},\phi_{0})$ is given by
\beq
\Tilde{F}(\zeta_{0},\phi_{0})=-\Omega_{\text{R}}\Omega^{2}\zeta_{0}\phi_{0} .
\eeq
The linearized equation \eqref{BJJLangevin} involves a harmonic potential and gives no quantum correction that stems from the higher-order derivatives of the external potential. 
The stationary solution is given by
\beq
\mathcal{P}_{\text{stat}}(\zeta_{0},\phi_{0})
=Z^{-1}\mathrm{exp}\left[-\frac{\hbar\Omega}{2k_{\mathrm{B}}T_{\mathrm{eff}}}\bar{N}\left(\frac{\Omega}{\Omega_{\text{R}}}\zeta_{0}^{2}+\frac{\Omega_{\text{R}}}{\Omega}\phi_{0}^{2}\right)\right] .
\label{BJJFPeqstat}
\eeq
We show the stationary solution in Fig.~\ref{PstatJJ}(b) for $\Omega/\Omega_{\text{R}}=0.5$ and $k_{\rm B}T/(\hbar\Omega)=0.1$. 
Figure \ref{PstatJJ}(b) shows that the relatively localized $\phi_{0}$ around the origin and the delocalized $\zeta_{0}$ are realized as a stationary configuration. 
As one decreases the ratio between the interaction energy and the tunneling energy $\Omega/\Omega_{\text{R}}=\sqrt{2g\bar{\rho}/J_{0}}$ less than one, the population imbalance $\zeta_{0}$ is delocalized and the relative phase $\phi_{0}$ is highly localized in the stationary configuration given by Eq.~\eqref{BJJFPeqstat}.

\section{Conclusions} 

In the first part of this paper, we have derived the 
Schwinger-Keldysh action of a particle under 
the effect of a deterministic external potential and 
a stochastic Ohmic bath, which contains both thermal and 
quantum fluctuations. 
Contrary to previous papers \cite{schwinger,keldysh,kamenev}, our derivation has been 
performed starting from the Langevin equation of the system. 
In the second part of the paper, we have then adopted the Schwinger-Keldysh action to include the velocity of the particle by using a Hubbard-Stratonovich transformation and to derive the fully analytical semiclassical and quantum Fokker-Planck equations for the time-dependent probability of the particle in the quantum-thermal Ohmic bath. 
The semiclassical Fokker-Planck equation involves the effective temperature associated with the frequency of harmonic potential. 
The obtained results can be applied to various contexts. 
In Sec.~\ref{SecFPJJ}, we wrote down the quantum Fokker-Planck 
equations for the Josephson mode in an atomic Josephson junction and for the superconducting phase in a superconducting Josephson circuit. 
We showed the stationary solution of the semiclassical Fokker-Planck equation for each of the Josephson systems. 
These Josephson systems have been experimentally realized and attracted marked attention. 
We expect that our work would also contribute to the understanding of such a noisy Josephson junction. 
For instance, the escape rate of the superconducting phase from a local potential minimum is related to the temperature that appeared in the Fokker-Planck equation \cite{kamenev}. 
Experimental measurements imply that the escape temperature deviates from the absolute temperature at a low-temperature regime. 
The deviation of the escape temperature is explained by macroscopic quantum tunneling \cite{clarke,clarke1988,blackburn}. 
In addition to the macroscopic quantum tunneling, within the underdamped limit, our obtained effective temperature originating from the quantum fluctuations would give a considerable contribution to this escape temperature in a superconducting Josephson circuit, which would be useful to verify our result. 
To obtain a quantum Fokker-Planck equation without higher-order derivatives, it could be useful to apply the approach of effective action \cite{helias}. 
The effective action includes the quantum fluctuations, and it would enable us to derive a quantum Fokker-Planck equation. 
\newline

\noindent
{\bf Acknowledgments}

\noindent
The authors acknowledge Pietro Faccioli, Amos Maritan, Fabio Sattin, and Sandro Azaele for useful suggestions. 
\newline 

\noindent
{\bf Authors' contributions} 

\noindent
KF and LS equally contributed to all aspects of the manuscript. 
Both authors read and approved the final manuscript.
\newline 

\noindent
{\bf Availability of data and materials}

\noindent
The data generated during the current study are available from the contributing author upon reasonable request.
\newline 

\section*{Declarations}

\noindent
{\bf Ethical approval and consent to participants} 

\noindent
The authors declare they have upheld the integrity of the scientific record.
\newline 

\noindent
{\bf Consent for publication}  

\noindent
The authors give their consent for publication of this article.
\newline 

\noindent
{\bf Competing interests} 

\noindent
The authors declare no competing interests. 
\newline 

\noindent
{\bf Funding}

\noindent
KF is supported by a PhD fellowship of the Fondazione Cassa di Risparmio di Padova e Rovigo. 
\newline

\noindent
{\bf Author details} 

\noindent
$^{1}$ Dipartimento di Fisica e Astronomia "Galileo Galilei" and QTech Center, 
Universit{\`a} di Padova and INFN Sezione di Padova, 
Via Marzolo 8, 35131, Padova, Italy. 

\noindent
$^{2}$ Istituto Nazionale di Ottica del Consiglio Nazionale delle Ricerche,  
Via Nello Carrara 2, 50019, Sesto Fiorentino, Italy.

\begin{appendices}
\section{Derivation of Fokker-Planck equation}\label{AppA}

In order to derive the semiclassical Fokker-Planck equation for $\mathcal{P}(q,v,t)$ 
from Eqs.~\eqref{fis07} and \eqref{fis03}, 
let us consider an infinitesimal time interval $\varepsilon=t'-t$. 
The probability $\mathcal{P}(q',v',t+\varepsilon)$ satisfies, 
to $\order{\varepsilon}$ \cite{das,kleinert},
\beq
\bal
&\mathcal{P}(q',v',t+\varepsilon) 
=\int^{\infty}_{-\infty} dq
\int^{\infty}_{-\infty} dv 
K(q',v',t+\varepsilon|q,v,t)\mathcal{P}(q,v,t)\\
&= \int^{\infty}_{-\infty} dq\int^{\infty}_{-\infty} dv 
\int_{q(t)=q}^{q(t')=q'} 
D[q(\bar{t})]\int_{v(t)=v}^{v(t')=v'}D[v(\bar{t})]\int 
D[\Tilde{q}(\bar{t}),\lambda(\bar{t})] \\
&\times\exp\left[\int^{t+\varepsilon}_{t}dt_{1}i
\Big[-m\Tilde{q}(t_{1})\dot{v}(t_{1})-m\Tilde{q}(t_{1})
\frac{\partial F(q,v)}{\partial v}
+\frac{i}{2}\int^{t+\varepsilon}_{t} 
dt_{2}\Tilde{q}(t_{1})C(t_{1}-t_{2})\Tilde{q}(t_{2}) 
+\lambda(t_{1})\left[\dot{q}(t_{1})
-v(t_{1})\right]\Big]\right] .
\eal
\label{Prob0}
\eeq
Taking into account Eq.~\eqref{whiteapp}, Eq.~\eqref{Prob0} gives
\beq
\bal
\mathcal{P}(q',v',t+\varepsilon)
&=\int^{\infty}_{-\infty} dq\int^{\infty}_{-\infty} dv 
\int D[\Tilde{q}(t),\lambda(t)] \\
&\times\exp\left[-i\varepsilon\Tilde{q}m\left(\frac{v'-v}{\varepsilon}
+\frac{\partial F}{\partial v}\right)
-\varepsilon\gamma 
k_{\mathrm{B}}T_{\rm eff}\Tilde{q}^{2}
+\order{\varepsilon^{2}}
+i\varepsilon\lambda\left(\frac{q'-q}{\varepsilon}-v\right)\right] 
\mathcal{P}(q,v,t)  \\
&=\int^{\infty}_{-\infty}dq\int^{\infty}_{-\infty}dv\int D[\Tilde{q}(t)]
\delta\left(q'-q-\varepsilon v\right)e^{-im\Tilde{q}(v'-v)} \\
&\times\left[1-i\varepsilon\Tilde{q}\left(m\frac{\partial F}{\partial v}-
i\gamma 
k_{\mathrm{B}}T_{\rm eff}\Tilde{q}\right)+\order{\varepsilon^{2}}\right]
\mathcal{P}(q,v,t) ,
\eal
\label{Prob1}
\eeq
In the above calculation, we used $\int D[\lambda(t)]
e^{i\lambda(t)\left[q'(t)-q(t)-\varepsilon v(t)\right]/\hbar}=\delta\left[q'(t)-q(t)
-\varepsilon v(t)\right]$. 
Hence, performing the integrals in Eq.~\eqref{Prob1}, one obtains
\beq
\bal
\mathcal{P}(q',v',t+\varepsilon)
&=\left[1+\varepsilon\partial_{v'}
\left(\frac{\partial F}{\partial v'}+\frac{\gamma}{m^{2}}
k_{\mathrm{B}}T_{\rm eff}\partial_{v'}\right)+\order{\varepsilon^{2}}\right]
\mathcal{P}(q'-v'\varepsilon,v',t) \\
&=\left[1+\varepsilon\partial_{v'}\left(\frac{\partial F}{\partial v'}
+\frac{\gamma}{m^{2}}
k_{\mathrm{B}}T_{\rm eff} \partial_{v'}\right)+\order{\varepsilon^{2}}\right]
\left(1-\varepsilon v'\partial_{q'}\right)\mathcal{P}(q',v',t) \\
&=\left[1+\varepsilon\left[-v'\partial_{q'}+\partial_{v'}
\left(\frac{\partial F}{\partial v'}+\frac{\gamma}{m^{2}}
k_{\mathrm{B}}T_{\rm eff} \partial_{v'}\right)\right] 
+\order{\varepsilon^{2}}\right]
\mathcal{P}(q',v',t) \; . 
\eal
\eeq
For $\varepsilon\to +0$ one finally finds Eq.~\eqref{FPeq}, which is our semiclassical Fokker-Planck equation. 
\end{appendices}

\end{document}